# Microprocessor Design with Dynamic Clock Source and Multi-Width Instructions

Keyu Chen*, *Student Member, IEEE*, Xuyi Hu*, *Student Member, IEEE*, and Robert Killey, *Fellow, IEEE*

***Abstract*—This paper introduces a novel 32-bit microprocessor, based on the RISC-V instruction set architecture, is designed,utilising a dynamic clock source to achieve high efficiency, overcoming the limitations of hardware delays. In addition, the microprocessor is also aimed to operate with both base (32-bit) instructions and 16-bit compressed instructions. The testing of the design is carried out using ModelSim with an ideal result.**

***Index Terms*—Microprocessor, reduced instruction set computer (RISC), dynamic clock source, compressed instructions, RISC-V instruction set architecture, SystemVerilog**

## I. Introduction

MOST current microprocessors, including both complex instruction set computer (CISC) and reduced instruction set computer (RISC) types[1], are driven by fixed-frequency clock sources. However, the minimum clock period, and hence maximum processing speed, is limited by hardware delays. In order to ensure successful operation, the clock period should be longer than the time required by the longest signal path in the circuit (referred to as the *critical path*), and this requirement results in significant time wastage, as the signal paths for the majority of the instructions are significantly shorter, and hence would function correctly for shorter clock periods. In single cycle microprocessors (as opposed to pipelined implementations), this issue is more serious since all possible delays within the execution of an instruction are summed together within a single clock period. Another issue is the waste of instruction memory space. For example, in the RISC-V instruction set architecture, the data width dedicated for immediate values within the instruction set is 12 bits which could represent a number from 0 to 4095 (or from -2048 to +2047 in the case of two's complement signed numbers. Such a range is too large in many applications.

Due to limitations of hardware devices, instructions have delays which vary over the range 0.4ns to 1.6ns. The respective delay for individual instructions depends on their critical paths; normally access to data memory or multiplexers will cause larger delay. The clock frequency is fixed in traditional microprocessors; thus, the clock period should be longer than the longest delay within the instruction set. However, this means that the microprocessor has variable periods of time when no instruction operation is being performed, an inefficient use of time and energy. Jia, Tianyu et al. add an extra clock management which is a multi-phase all digital phase locked loop. [2] This technique can dynamically change the period of the clock by selecting different phases from the phase locked loop without varying the original frequency of the clock source. Researchers choose bits [31:28] in each instruction to represent the 16 cases of phase. The clock period could be varied between +40% and -30% of the original frequency. By implementing a 55nm low-power CMOS chip in extra clock management, test results show that the processing speed rose by 18% and energy efficiency rose by 22%. [2] Power efficiency of microprocessor has been improved successfully by applying this technique. The theory used is also easy to understand and apply. However, the architecture used is a 5-stage pipeline structure which increases the complexity, since additional circuits are needed to solve the resulting hazards. For example, in order to ensure the correct values are held within the pipeline registers during the process of phase changing, a calibration device is applied. This device has an additional dedicated register which is driven by a conventional clock source. An XOR gate compares these two registers and generates an error

This paragraph of the first footnote will contain the date on which you submitted your paper for review, which is populated by IEEE. It is IEEE style to display support information, including sponsor and financial support acknowledgment, here and not in an acknowledgment section at the end of the article. For example, "This work was supported in part by the U.S. Department of Commerce under Grant BS123456." The name of the corresponding author appears after the financial information, e.g. *(Corresponding author: M. Smith).* Here you may also indicate if authors contributed equally or if there are co-first authors.

Keyu Chen is with Department of Electrical Engineering, University College London, London, WC1E 6BT UK (e-mail: zceehen@ucl.ac.uk).

Xuyi Hu is with Department of Computing, Imperial College London, London, SW7 2BX UK (e-mail: xh519@ic.ac.uk).

R. I. Killey is with the Optical Networks Group, Department of Electronic and Electrical Engineering, University College London, Torrington Place, London WC1E 7JE, UK (e-mail: r.killey@ee.ucl.ac.uk).

signal when these two values are different. This process costs more energy which can be avoided. Meanwhile it means multiple instructions with different delays need to be processed at the same time; thus, the energy saving in the pipelined case will not be as high as for single cycle architecture. However, the concept of exploiting the differences in instruction delays is quite useful. A dynamic clock is certainly an ideal solution to improve utilization of clock period. The method introduces by Jia, Tianyu et al is achieved using the ARM instruction set architecture, in which the formats of instructions are also edited sightly.

A new method is introduced to improve power efficiency by replacing the conventional RISC architecture by a stack architecture. Waterman, Andrew et al. highlights that standard RISC architectures have low power efficiency due to their complexity. [3] Multiplexers are used to connect register files to functional units, and the complexity of this system will increase with more registers and functional units. Application of a pipeline structure is also increasing the complexity; all these hardware designs increase the number of transistors and thus lead to a rise in the power consumption. In comparison, a stack machine has a lower number of transistors and is easier to decode since instructions and data are always extracted from the top of the stack. Although this design will undoubtedly increase the power efficiency, it is decided not to use this idea. Since an entire new architecture means a new instruction set and corresponding expansion package of instruction set, it will certainly be definitely a large amount of work. Estimated improvement in efficiency of this technique is approximately 10-15%, and it is decided that this level of optimization is not worth such a high level of work. It will be more efficient to optimize the original architecture and instruction set of RISC-V. [3]

## II. Typical design of microprocessor

The microprocessor designs in this project is based on the RISC-V instruction set architecture with limited extension of compressed instructions. Single cycle structure is suitable for initial design of a new technology due to its low complexity [4]. RISC-V[5] is the most suitable ISA for this project since it is totally open sourced. Harvard architecture[6] is applied to this design to separate instruction memory and data memory.

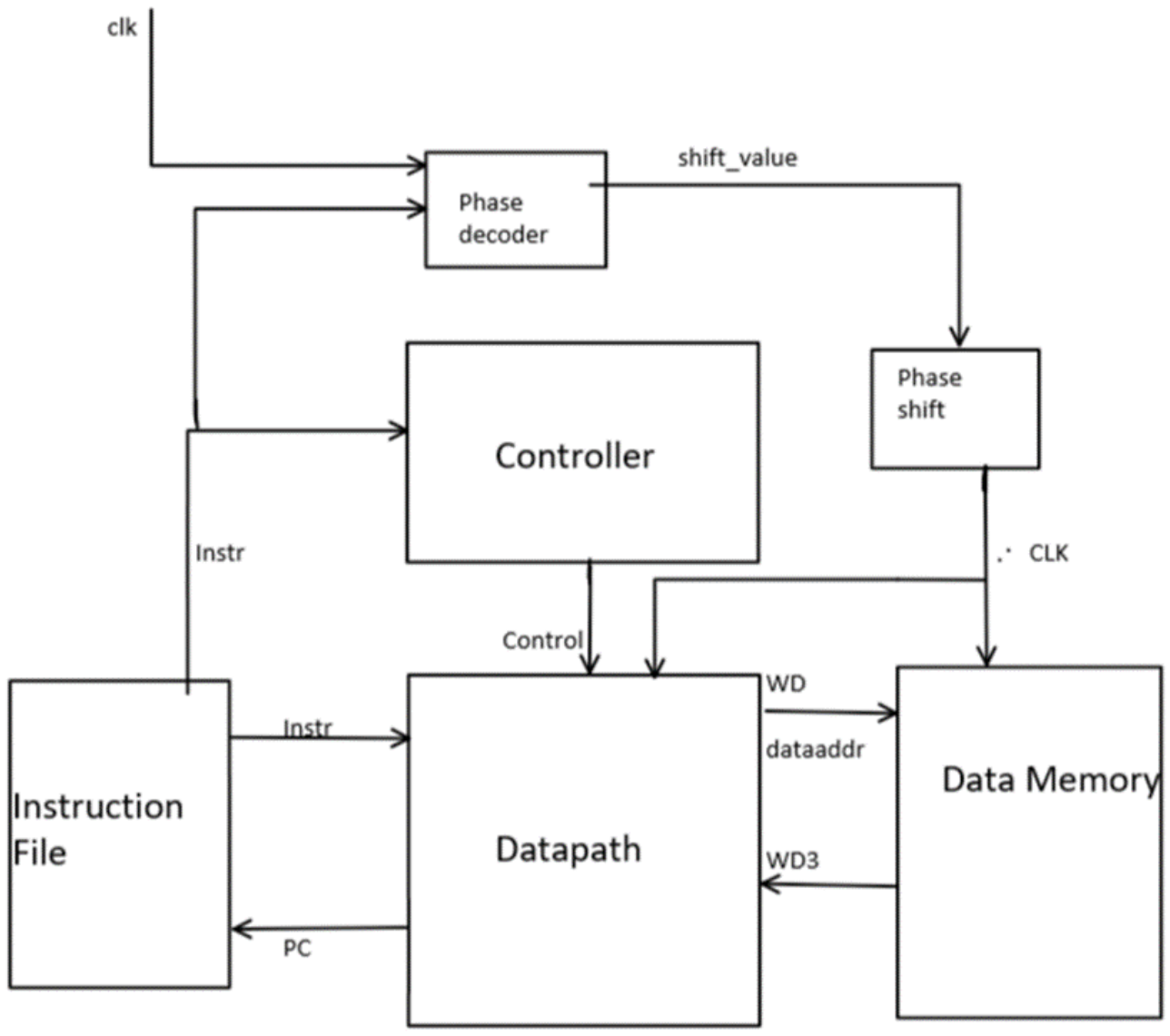

Figure 1. Block diagram of entire microprocessor.

The phase decoder is used to decode the instructions stored in instruction file and allocate them the corresponding shift value, which will be sent to the phase shifter which will extend or reduce the clock period which will be sent to the microprocessor. The control signal is a combination of multiple signals applied to multiple units within the data path such as the register file, the arithmetic and logic unit (ALU) and multiplexers. The instruction file should contain both 32-bit instructions and 16-bit instructions, so the data path will be designed to operate with both kinds of data width.

This clock source is dynamic, which means that its clock period varies based on the type of instructions. This clock source needs a master clock to give a basic clock signal and it will be processed by a phase shifter to produce a new clock signal for microprocessor.

Similar to the decoder in the controller circuit, the phase decoder uses the same method to identify the type of instructions. The system obtains a phase shift value based on delays of different instructions.

In actual hardware devices, delays are unavoidable, it can be caused by hardware limits of signal transfer through the microprocessor circuit. In a traditional RISC microprocessor, delays of register file, ALU and data memory are much higher than any other functional units [7]. Thus, I apply read delay to register file, transferring delay to ALU and read delay to data memory. Different instructions have different critical paths; thus, they have different values of delay. For now, the longest delay in this project is 16ns. Among base instructions of RISC-V, LW has the longest delay due to longest critical path, this feature is observed with actual microprocessors.[7] Thus, the phase decoder should allocate longest phase shift value for this instruction.

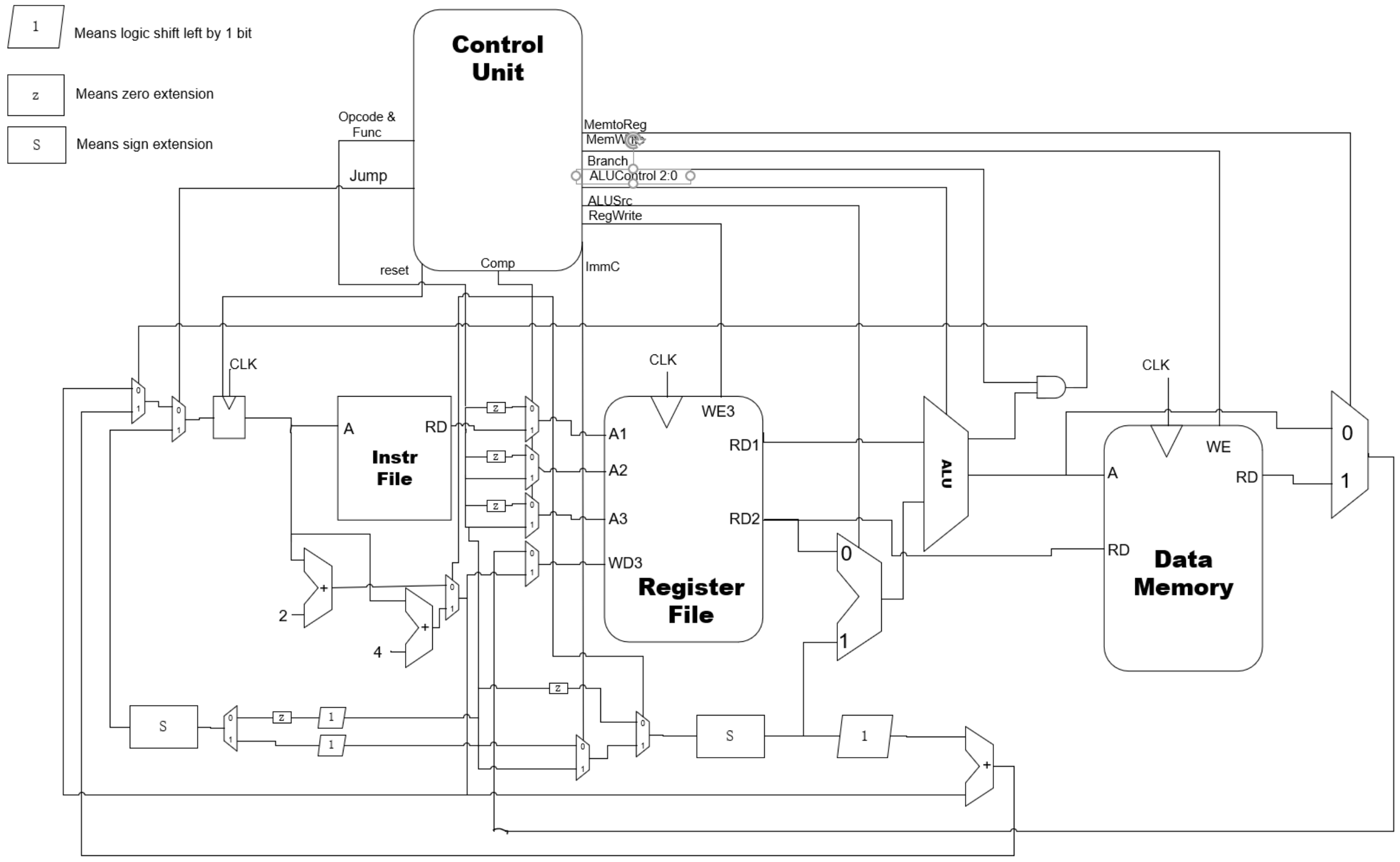

Figure 2. Overview of Hardware Circuit.

Figure 3 indicates the SystemVerilog code for phase decoder.

The phase decoder examines the opcode instructions to identify the type of them. The shift value given by the phase decoder means the scale of extension of clock period, the actual clock period obtained should be multiples of the master clock period, and the value of multiple equals to phase shift value. Since LW has the longest delay, phase decoder allocates it largest shift value which is 8. Compressed version of instructions has the same critical paths so their shift value allocated will not be changed. The current module generate 3 stages of shift value since there are 3 kinds of delay in the project, A larger number of shift values could be implemented if more complex delays are added to the microprocessor.

The clock source should work correctly without the input from instructions, thus, another input “reset” is added to the phase decoder. Once this signal is enabled, the system will be given a fixed shift value until a new input is applied from the instruction memory.

The phase shifter which is shown in Figure 4 is intended to extend the clock period based on the shift value given by phase decoder. In this module, the function of phase shift is achieved by a counter.

The counter increments at each rising edge of clock source. The shift value is a threshold of the counter, the counted value will drop to 0 once it reaches the threshold. The output is 1 when the counter is smaller than half of threshold, otherwise the output is 0. Clock signals generates by this method have relatively longer negative side. The master clock source is 500MHz in this case, thus the actual clock period generated is 18ns if the shift value is 8. This method creates a redundant backup to ensure successful command execution.

| Instruction | Opcode | ImmC | RegWrite | ALUSrc | Branch | MemWrite | MemtoReg | Jump |
|---|---|---|---|---|---|---|---|---|
| R-type | 0110011 | X | 1 | 0 | 0 | 0 | 0 | 0 |
| LOAD | 0000011 | 1 | 1 | 1 | 0 | 0 | 0 | 0 |
| STORE | 0100011 | 0 | 0 | 1 | 0 | 1 | X | 0 |
| BEQ | 1100011 | 0 | 0 | 0 | 1 | 0 | X | 0 |
| JAL | 1101111 | X | 1 | X | X | 0 | X | 1 |
| ADDI | 0010011 | 1 | 1 | 0 | 0 | 0 | 0 | 0 |

Table 1. Truth table of the control unit.

As with the PC register and phase decoder, the phase shifter could reset its count value. The phase decoder generates 3 shift values which are 2, 6 and 8, thus the clock source should be able to generate clock periods with 6ns, 14ns and 18ns.

The rest of microprocessor does not have obvious difference with traditional RISC architecture, but it still needs to be optimized to apply compressed instruction set. The hardware design is based on the instruction list given in the RISC-V instruction set manual.[5]

The main circuit of the microprocessor which is shown in Figure 2 is divided into 3 parts, memory units, data path and control unit. The memory units include data memory and instruction memory, they are required to store operation parameters and instructions respectively. Theoretically, the instruction memory should be byte addressing in order to apply multiple width instructions. For the same reason, the program counter is set to 2 modes of increasing address naturally: +2 and +4.

Data path includes most of functional units of the microprocessor which execute instructions. In order to match 16-bit instructions, multiple multiplexers and extension units are added to the data path, this will be explained later in this section. The bit width and bit order of each typical connecting wire depends on the typical format of instructions they are designed for.

The final part of the hardware design is the control unit. The complete diagram is shown above. The control unit is used to control the operation of the entire architecture. It has two inputs: Opcode and Funct3, and their positions in instructions are fixed. Opcode has 7 bits which control all multiplexers, data memory and register file and funct3 assists the control of the ALU. All R-type instructions have the same Opcode which is 0110011, and the instruction specifies two source registers and one destination register. An immediate number is not required, so the output of ImmC is X (i.e., 'don't care'). The result needs to be written back to the register file so RegWrite needs to be 1. Both RD1 and RD2 are input to the ALU, so ALUSrc is 0. Branching is not carried out, so Branch is 0, Jump is 0. Data is not being stored in data memory so MemWrite is 0, while data from memory is not needed, so MemtoReg is 0.

The R-type instruction has another input to the control unit, which is funct7, it is used with funct3 to assist to control the ALU. Most Funct7 of R-type instructions are 0000000, only SUB has 0100000.Same with R-type instructions, we can derive the truth table for other instructions which are shown in Table 1.

The compressed instruction set was included the hardware architecture. The compressed instruction has 16 bits which means it could save space and energy. The entire system needs a control signal from control unit to switch to compressed instruction mode when necessary. The key for the control unit to distinguish between base (32-bit) instruction set and compressed (16-bit) instruction set is the bit 0 and bit 1. Those two bits are always 1 in compressed instructions and will never be both 1 in base instructions. Thus, if the first two bits of Opcode are not both 1, the control unit will generate a control signal Comp (Comp=1 means the system is executing a normal 32-bit instruction). Compressed instructions have different formats. Multiplexers controlled by Comp must be added before the input of the register file and sign extension to select appropriate inputs. Zero extension (square box) is applied before the multiplexer to expand the data to the same length as used in the base instructions. The program counter needs to be edited to support 16-bit instructions. The address of next instruction should be obtained by adding 2 instead of 4 when executing compressed instructions (since compressed instructions are 2-byes wide, rather than 4-bytes). For the same reason, units design for both conditional and unconditional branch need to be optimized. Immediate bits should be shifted left by 1 instead of 2 when executing compressed instructions, and zero extension is indispensable. The remaining part of the architecture does not need to change, since the difference only appears in the decoding stage.

The current design cannot support many compressed instructions, because they are not as ordered as base instructions. More units need to be added if more instructions are supported. In my opinion, increasing the complexity to support a low number of compressed instructions is not valuable; thus, I only made limited changes to my design to support a limited number of compressed instructions. For now, this design is available for most of compressed version of instructions mentioned in base design, except ADDI since the position of the immediate number is different from the rest of the instructions.

## III. MODELSIM SIMULATION

The entire design was simulated with ModelSim. All functional units described by Hardware Description Language (HDL) modules, and SystemVerilog[8] was chosen for this project. This section will describe SystemVerilog code for several critical functional units.

The register file is a set of registers. In base RISC-V ISA, the register file has 32 registers, each register has 32-bits. The addresses of registers are listed from 0 to 31 in order. Register0 is set to be 0 eternally. The register file has 6 inputs: A1, A2, A3, WD3, WE3, clk. A1 and A2 will read the input value which should be the addresses of registers, and two outputs, RD1 and RD2 will output the values stored in both registers respectively. If WE3 is 1, the register file is allowed to edit the register file, in this case, A3 will give the address of register which is going to be written to, and WD3 is the value to be written.

Figure 5 describes the register file. The data width of A1, A2 and A3 are set to 5 bits to match RISC-V instruction format. The memory space of this module is set to 32 * 32 to simulate a register file. Data is only allowed to be written into the register file at the rising edge of the clock when WE3 is enabled. Outputs RD1 and RD2 do not have such limitations, ideally they are able to give the output once correct addresses are sent to A1 and A2. In this case I added 6ns delay to this process to simulate the actual delay due to hardware. Please be aware that this value is only used for simulation, the actual value for delay could be quite different with 6ns.

The ALU which is described in Figure 6 is another critical functional unit of the data path. It applies multiple logic or arithmetic operations to the data input. The type of operation is controlled by another input signal ALUControl. The output of ALU is the result of operation. In this case, first input of ALU (SrcA) comes from RD1 of register file, and SrcB need to be selected between RD2 and immediate value. The complexity of wires within ALU is highly affected by the number of operations which the ALU supports, thus, the hardware delay of this part cannot be ignored. Current ALU has 8 logical or arithmetic operations which are controlled by input signal “ALUControl”. “AND”, “OR”, and “XOR” are all basic logic operations, addition and subtraction are both normal arithmetic operations. SLL mean logical shift left and SRL means logical shift right, in both operations SrcA is the value to be shifted and SrcB is the scale of shift.Logical shift ignores the effect of sign bit, if sign bit is eliminated due to shift operation, the system will not make any compensations is slightly different with other operations, it compares inputs from both SrcA and SrcB, the output will be 1 if SrcA is larger than SrcB, otherwise the output should be 0. Note that this module has one more output signal which is called zero, it is a unique output for subtraction. As the instruction “BEQ” asks the ALU to execute subtraction, once the result is zero, output “zero” will generate a new signal. Signal “zero” should be 0 or x in any other cases.

After defining all necessary modules, a new module called "data path" needs to be defined to combine all functional units together, please see the full code in Appendix. The data path receives input from both data memory and instruction memory. The controller sends 8 kinds of control signal to data path. All of the signals which leave or reach the data path are defined by “input logic” or “output logic”. Other variables which are defined as “logic” are wires within the data path.

Building the design block by block is much easier to understand the entire structure. The data path was subdivided into 3 blocks: program counter (or “next pc”), register file and ALU.

Block “next pc” includes all functional units required for the program counter; this block should be able to provide correct instruction address. This program counter needs a register, several adders, multiplexers and extensions. Second and third rows of this block refers to two adders, which add 2 or 4 to the instruction address, respectively. The instruction memory is fixed at 32 bits. 16-bit compressed instructions will be expanded to 32-bit automatically, thus the parameters for these two adders are edited. In addition to the usual 2- or 4-increase of the instruction address, the program counter also accepts extra address inputs from jump instructions. Zero extension and multiplexers are required to switch between base instructions and compressed instructions. The data width of the immediate number in the compressed instructions needs to be expanded before passing through the multiplexers since both inputs of both multiplexers should have same data width. Due to the format of the RISC-V instruction set architecture, some bits of instructions given by the instruction file are reordered (this is implemented in the HDL using curly brackets). Block “register file” includes all functional units for a register file. This block should give an output based on the address input, and store the input data at the rising edge of the clock when the write-enable signal is asserted. This Block has far fewer elements than the program counter. Extra multiplexers and zero-extension units are added before the register file input due to mixed use of 32-bit and 16-bit instructions. The last block is the ALU; it includes an ALU module and a multiplexer to select the second input of the ALU, signal source of SrcB could be either RD2 of the register file or the immediate value after sign extension.

| 0 | ADDI | R0 | | 5 | R1 | 00000000010100000000000010010011 | 5 | X |
|---|---|---|---|---|---|---|---|---|
| 4 | ADDI | R0 | | 7 | R2 | 00000000011100000000000100010011 | 7 | X |
| 8 | ADD | R1 | R2 | | R3 | 00000000001000001000000110110011 | 12 | 7 |
| 12 | ADDI | R0 | | 5 | R4 | 00000000010100000000001000010011 | 5 | X |
| 16 | SW | R0 | R4 | 0 | | 00000000010000000010000000100011 | 0 | 5 |
| 20 | LW | R0 | | 0 | R5 | 00000000000000000010001010000011 | 0 | X |
| 24 | SUB | R2 | R4 | | R6 | 01000000010000010000001100110011 | 2 | 5 |
| 28 | C.AND | R2 | R3 | | R2 | 1000110101101101 | 4 | 12 |
| 32 | ADD | R2 | R1 | | R7 | 00000000000100010000001110110011 | 9 | 5 |
| 36 | JAL | | | 44 | R8 | 00000010110000000000010001101111 | X | X |
| 40 | ADD | R7 | R0 | | R9 | 00000000000000111000010010110011 | | |
| 44 | ADD | R8 | R1 | | R10 | 0000000000101000000010100110011 | 40 | 5 |

Table 2. Instruction list for test bench**.**

Figure 7 shows the code for instruction memory. SystemVerilog code for both instruction memory and data memory are similar, both of them are set to word aligned, the data memory is set 4ns delay for appropriate simulation.

Controller circuit is an indispensable part of a microprocessor. It receives instructions from instruction memory and sends control signals to both data path and data memory. The controller circuit checks the opcode, func3 and func7 to identify the type of instructions and provides corresponding control signals, as explained in Section 1.4. First step of controller unit is examining first two bits of the instruction, once both of it are 1, the output of signal “comp” is 1, otherwise it is always 0. In case of base instructions, the controller will then examine the first 7 bits of the instruction which is called the opcode. Controller circuit will carry out further examination of func3 and func7 when executing R-type instructions. The operation on compressed instructions is generally the same, compressed instructions also have opcode, func3 and func7 although their data widths have been reduced. The current controller circuit supports 8 kinds of R-type instructions, load word, save word, addi (add immediate), BEQ, jump and link. Compressed instructions including 4 kinds of R-type instructions, load word, save word, jump and link are executable within this microprocessor. Note that all functions described above are achieved by a decoder instead of controller; the decoder is a sub-unit of the controller, a complete controller needs one more AND gate to execute BEQ. That is why this SystemVerilog module(see Appendix) is named decoder.

After finishing the functional units, we need a top design to integrate them. The first hierarchy should be able to combine data path and controller circuit to achieve initial ability of data processing. This module is called processor. The controller is a combination of decoder and an AND gate for signal “branch”. The processor needs to receive instructions from instruction file and data from data memory. Data memory is also controlled by the controller, thus, required control signal is defined as an output of processor.

The top design is a combination of all functional units within the microprocessor, including dynamic clock source, controller, data path, instruction memory and data memory.

The last step before simulation is creating an instruction file for instruction memory, the format of instructions is based on instruction lists within RISC-V instruction set manual.[5]

Many instructions need to call data stored in register file, however a new defined register has no data saved. Thus, first two instructions have to be addi so that data can be saved to register file for future use.

In order to prove unique designs for compressed instructions, instruction list should contain several 16 bit compressed instructions.

Table 2 is the instruction file used for final test. First two instructions are used to assign register 1 and register 2, the third instruction is used to test R-type instructions. Load and save should be achieved by the next two instructions. The 8th instruction is a compressed instruction to prove availability of functional units designed for compressed instructions, compressed instructions share the same truth table with base instructions so it is not necessary to execute too many compressed instructions. The 10th instruction is an unconditional jump instruction and its target is 8 bytes after its own address, thus the instruction after it should not be executed. The last instruction is used to check if the address of JAL is written into the register file successfully. The last two columns of the file are expected outputs of the test, X means “do not care” so the actual output can be filled with 0 or x. Details of simulation result will be discussed in next section.

The final simulation results should be obtained by running the final test bench.[9] After fixing all existing bugs, the final wave form should look like this:

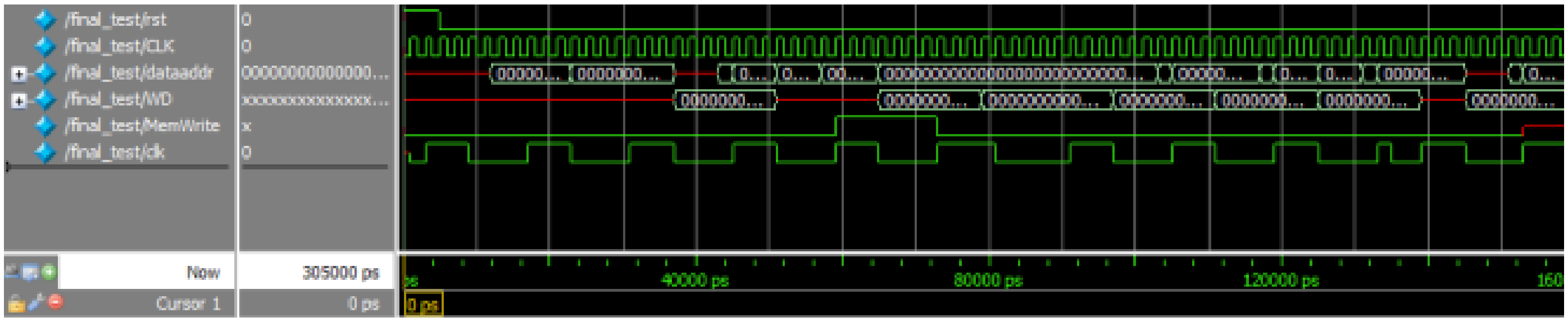

Figure 9. Wave form of first test.

Within this graph, CLK is the master clock source which is 500MHz, each division of the graph represents 10ns, clk is the output of dynamic clock source, “dataaddr” is the output of the ALU. Obviously, most of clock periods are 14 ns, because most of instructions used are R-type instructions or ADDI which have the same delay. The 6th clock period is longer than other clock periods because the 6th instruction is LW which does have longer delay than any other instructions. The 10th clock period is extremely short, because the 10th instruction is JAL which does not need to pass data path, it has shortest delay. Outputs have many glitches, there are multiple values of output in a single clock period. Glitches are obtained by delays, output of ALU may be affected by parameters from previous clock period. The output should be correct at the end of its clock period.

The first clock needs to add 5 to register 1, so the output of dataaddr should be 5. For the same reason, register 2 should be 7. The 3rd register needs to add values saved in register 1 and register 2, the output of ALU should be 12. ALU output for 4th clock period should be 5. The output of SW should be the sum of register 0 and immediate 0, the result is also 0. LW is same with SW, output is 0. The next instruction is subtraction between register 2 and register 4, the ALU output should be 2. The 8th instruction should be 4 since it applies AND operation between 7 and 12, thus the result is 4. The result for 9th instruction should be 9 instead of 12 because the 8th instruction wrote the result back to register 2. JAL should write current address to register instead of read it, thus we do not expect any output. The next clock period is prepared for 12th instruction since 11th instruction is skipped, thus the ALU output is the sum of 40 and 5 which is 45.

It may not be that convenient to check ALU output by viewing wave plots since the time window for correct output is too small. Another method to prove successful operation is checking memory space of data memory and register file.

For R-type instructions and ADDI, output of ALU should be written back to destination register (rd), thus, register 1 is 5, register 2 is 2, register 3 is 12, register 4 is 5, register 6 is 2, register 7 is 9, register 10 is 45. SW and LW save value of register 4 to data memory and load back to register 5, thus register 5 should be 5. JAL save its current address to register 8, thus it should be 40

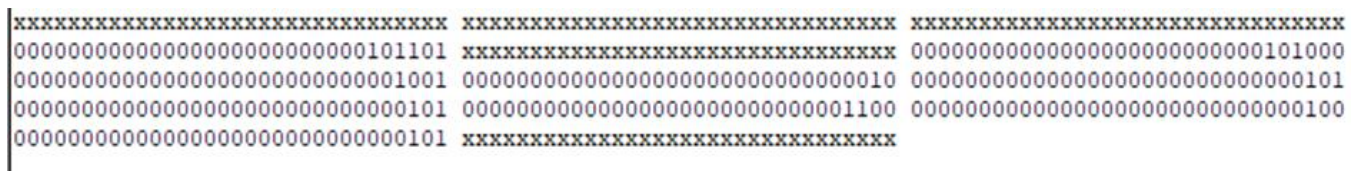

Figure 10. Register Memory of the module.

The register memory is shown in Table 3, clearly all registers hold same values with our expectation. It is also necessary to view data memory to ensure data is saved correctly.

It is essential to test the microprocessor with a program written by senior programming language to prove its availability and advantages. The program, should contain multiple J-type instructions to produce remarkable reduction of executing time.

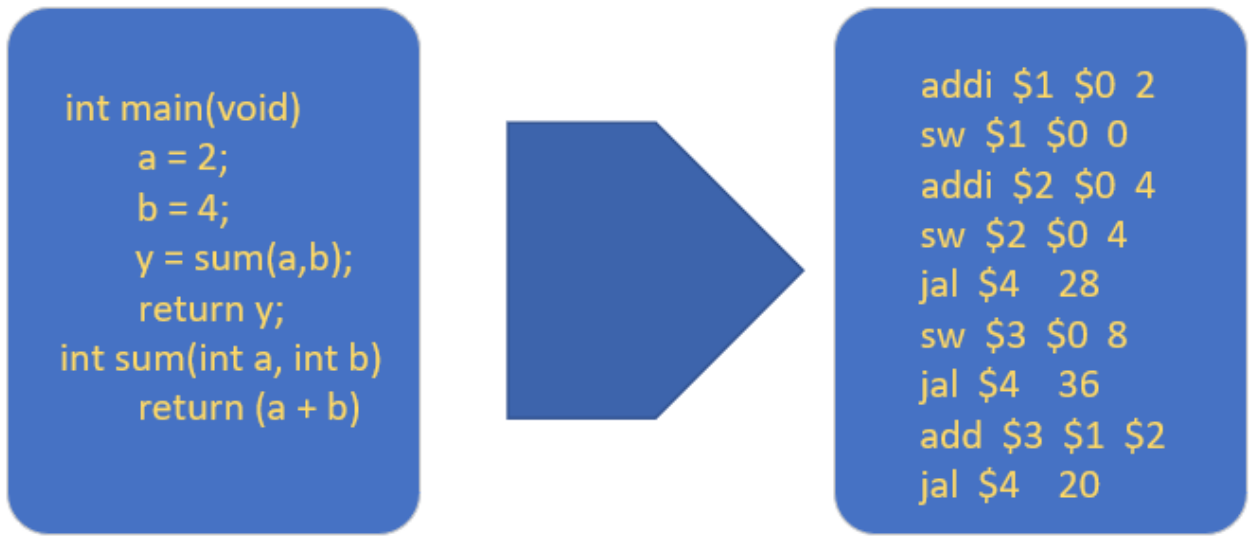

Figure 11. Program for testing and corresponding assemble language.

The test program is written in C language and be transferred to assembly language for test, the instruction file should contain 9 instructions and expect to execute within 162ns.

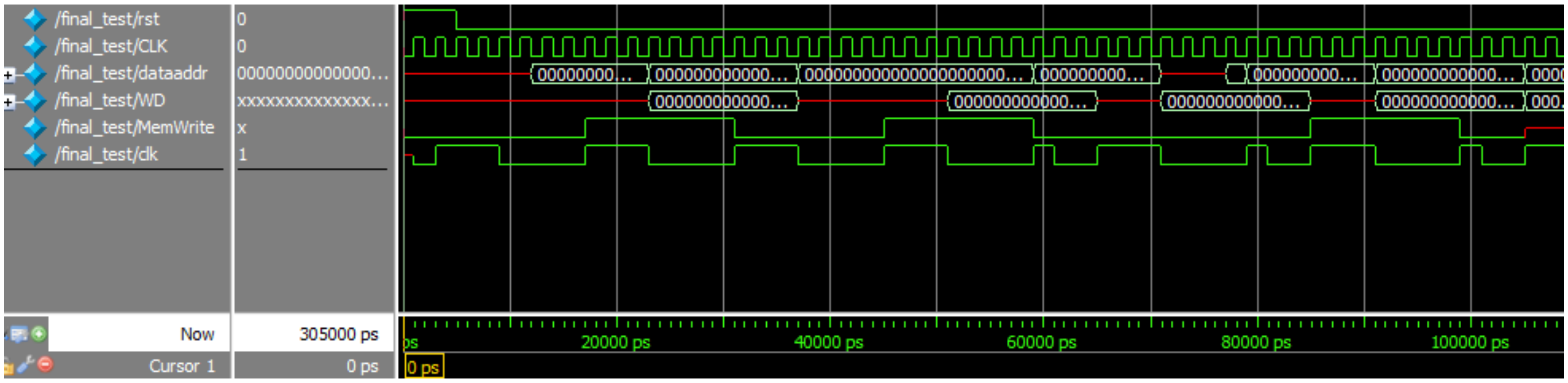

Figure 12. Waveform of second test.

Same with previous test, the simulation is successful, all data are written into correct positions, a remarkable reduction in operating time can be observed, executing 9 instructions take 105ns which is 35% shorter than traditional microprocessors, since one third instructions in the list are J-type which saves much time.

In conclusion, the wave form is in-line with our expectation, outputs are correct, clock period varies with the type of instructions, all data are written to corresponding register file and data memory correctly. Thus, this simulation is successful and the design of microprocessor with dynamic clock source is proved to be viable.

## IV. Conclusion

The simulation result of final test proves the viability of my design. The wave form given in the previous section shows that executing 10 instructions (ignore the instruction which is skipped), takes 155ns. However, executing the same program in the same microprocessor but without dynamic clock source takes 180ns since its clock periods must be at least 18ns, which is the longest possible delay for the full set of instructions. The efficiency of executing current instructions has been increased by 14% through the addition of a dynamic clock source. The efficiency will rise if higher proportion of instructions with short delay are executed since more short clock periods are utilized.

In addition, the machine code program tested included a compressed instruction and it was executed successfully. Thus, the simulation result also proves that this design is able to execute instructions with different data-widths, even though they are mixed together. More space in the instruction memory would be saved if the proportion of compressed instructions rises.

Although this design is successful, there are still many improvements to be made. This microprocessor is designed for proving the method of using a dynamic clock source.

Thus, the delay applied to it is simple, 13 instructions only have 3 separate values of delay, this is surely impossible in reality. Real hardware delays in actual microprocessors could be much more varied such as the delay of each multiplexer and the delay in writing into the register file. Besides, values of delay used in this project are used for test only, actual delays could be much shorter; if so, we need to prepare a new master clock source with higher frequency, and the phase decoder needs to set more stages of shift value. In a word, more accurate data about delays are needed before transferring the design to an actual microprocessor.

Another improvement that can be made is the complexity of microprocessor. For now, the microprocessor supports 13 base instructions and 7 compressed instructions. Its supported list of instructions could be much more if we add more functional units. Besides, current structure is single cycle which is the simplest structure. Similar with the paper described in the literature review [1], this technology could be tested on a pipeline structure although my instruction set architecture is extremely different to the one used in that paper. Pipelined technology is supposed to have larger throughput under same conditions, but the effect of dynamic clock source could be reduced seriously since an instruction needs multiple clock periods to be executed [10]. Delays for periods of decoding, fetching and executing [11] need to be considered separately, it increases the difficulty of predicting actual possible delay for each instruction and the effect of long delay instructions shall be increased.

Clock gating is an extra idea of this project, since many instructions such as R-type instructions do not need data memory, the phase decoder could generate an additional signal to cut the clock input to the data memory in order to save energy. This needs to be achieved by a switch controlled by that signal[12], unfortunately it cannot be simulated well with ModelSim, it should be able to test if the design is applied to real hardware devices

## VI. Appendix

```
module phase_decoder(input logic [6:0] opcode,
                input logic [4:0] op_c,
                input logic rst,
                output logic [3:0] shift_value);

always_comb

begin
  if (rst) shift_value<=4'b0110;
else
  begin
  if(opcode[1:0]==2'b11) // base instructions
    begin
      case(opcode)
        7'b0110011:shift_value = 4'b0110; //r-type
        7'b0100011:shift_value = 4'b0110; //SW
        7'b0000011:shift_value = 4'b1000; //LW
        7'b0010011:shift_value = 4'b0110; //addi
        7'b1100011:shift_value = 4'b0110; //beq
        7'b1101111:shift_value = 4'b0010; //jal
        default:shift_value = 4'b0110;
    endcase
    end
  else
    begin
      case(opcode)
        5'b10001:shift_value=4'b0110;
        5'b11000:shift_value=4'b0110;
        5'b01000:shift_value=4'b1000;
        5'b00101:shift_value=4'b0010;
        default:shift_value = 4'b0110;
     endcase
end
end
end

endmodule
```

Figure 3. SystemVerilog code for decoder.

```
module phase_shift(input logic CLK,
                input logic rst,
                input logic [3:0] shift_value,
                output logic clk);

logic [3:0] count;

always @ (posedge CLK)
begin
        if(rst)
          count<=0;
end


always @ (posedge CLK)
begin

        if(count<shift_value)
           count<=count+1'b1;
        else
           count<=0;
end


always @ (posedge CLK)
begin
        if(count<(shift_value>>1))
           clk<=1'b1;
        else
           clk<=1'b0;
end

endmodule
```

Figure 4. SystemVerilog code for phase shifter.

```
 `timescale 1ns/1ps
module register(input logic [4:0] A1, A2, A3,
 input logic [31:0] WD3,
 input logic clk, WE3,
 output logic [31:0] RD1, RD2);

logic [31:0] rf[31:0];


always_ff @ (posedge clk)
 if (WE3)
 rf[A3] <= WD3;

 assign #6ns RD1 = (A1!=0)?rf[A1]:0;
 assign #6ns RD2 = (A2!=0)?rf[A2]:0;
endmodule
```

Figure 5. SystemVerilog code for register.

```
module ALU(input logic [31:0] SrcA, SrcB,
        input logic [2:0] ALUControl,
        output logic [31:0] ALU_out,
        output logic zero);



always_comb

begin
  zero <= 0;
  case(ALUControl)
    3'b000 : ALU_out <= SrcA & SrcB; //and
    3'b001 : ALU_out <= SrcA | SrcB; //or
    3'b010 : ALU_out <= SrcA ^ SrcB; //xor
    3'b011 : ALU_out <= SrcA + SrcB; //add
    3'b100 : begin
        ALU_out <= SrcA - SrcB;
                begin
                  if(SrcA == SrcB)
                    zero <= 1;
                  else
                    zero <= 0;
                end
             end //subtraction
    3'b101 : begin
                if (SrcA <  SrcB)
                  ALU_out <= 32'd1;
                else
                  ALU_out <= 32'd0;
                end //slt
    3'b110 : ALU_out <= SrcA << SrcB[4:0]; //sll
    3'b111 : ALU_out <= SrcA >> SrcB[4:0]; //srl
    default: begin
             ALU_out <= 32'b0;
             zero <= 0;
             end
  endcase
end
```

Figure 6. SystemVerilog code for ALU.

```
module instruction_mem(input logic [31:0] addr,
                output logic [31:0] instr);

  logic [31:0] RAM[63:0];

  initial
        $readmemb("C:/intelFPGA_lite/20.1/instr.txt", RAM);

  assign instr=RAM[addr[31:2]];
endmodule
```

Figure 7. SystemVerilog code for Instruction Memory.